\documentclass{elsart}
\usepackage{epsfig}
\usepackage{natbib}

\def\url#1{#1} 

\begin{document}

\begin{frontmatter}

\title{Scattered starlight contamination in the spectrum of reddened stars.}

\author{Fr\'ed\'eric \snm Zagury}

\address{
\cty 02210 Saint R\'emy Blanzy, \cny France
\thanksref{email} }

  \thanks[email]{E-mail: fzagury@wanadoo.fr}

\received{May 2001}
\accepted{July 2000}
\communicated{L.L. Cowie}

 \begin{abstract}
The comparison, undertaken in preceding papers, of the UV observations of nebulae and of reddened stars 
reveals contradictory aspects of interstellar extinction.
The aim of this paper is to understand the implications hidden behind 
the apparent contradictions.
The questions treated will be: how can small grains with an isotropic 
phase function make an appreciable contribution in the UV spectrum of 
a star? Why are small grains not observed in the spectrum of a nebula?
How much of starlight can be scattered by large grains in the forward direction?        
\end{abstract} 

\end{frontmatter}

 \section{Introduction} \label{intro}
This paper will focus on the implications of manifest
contradictions which appear when UV spectra of nebulae  
(observations of a cloud at an angle $\theta>0$ from the illuminating 
star) and of reddened stars
(observations of a cloud at $\theta=0$) are compared.

The UV spectrum of a nebula, generally observed at close distance 
(within a few arcminute) from the illuminating star is the product of the 
spectrum of the star and a linear function of $1/\lambda$ (\citet{uv1}, UV1).
The grains responsible for the scattering must have a
strong forward scattering phase function (UV1).
The latter property point to large grains as the scatterers.
The UV extinction properties of these grains are very similar to what is observed in the optical (UV1).

The UV spectrum of reddened stars was decomposed into a direct 
starlight component and a component of light scattered at very small 
angle from the star (\citet{uv2}, UV2 and \citet{uv3}, UV3).
The scattered light component is due to grains with size small 
compared to the wavelength.
These grains, according to scattering theory, isotropically scatter 
starlight.
Excess of extinction does occur at $2200\,\rm\AA$ but the appearence of
the bump is due to the extinction of the scattered light, and not 
necessarily of the direct starlight.

Attention will focus on the following questions:
\begin{itemize}
    \item  How can small grains be so efficient at scattering light in 
    the forward direction (scattering angle $\varphi \sim 0$)?
Related questions such as the angular dependence of the scattering, the 
dependence of the scattering component with the distances star-cloud 
and observer-cloud will be discussed in this section.
    \item  Why is scattering by large grains a small fraction of the 
    direct starlight at $\varphi=0$? How much of the light is
    scattered by the large grains in the forward direction?
    \item  Why is there no bump in some nebulae?
    Why is the spectrum of a nebula the product of the 
    spectrum of the illuminating star and a linear function of 
    $1/\lambda$?
\end{itemize}¥
\section{Coherent scattering by small particles} \label{cosca}
\subsection{Scattering at $\theta=\varphi=0$ by small particles} 
\label{sca0}
The small grains which contribute to the UV spectrum of a 
reddened star have properties which are difficult to reconcile.

Because they are small compared to the wavelength, the scattering will 
be close to isotropic.
If so the scattered light should increase with the beam of the 
observation, but this is not observed: the scattering is efficient 
within a small angle only, smaller than $1''$, the
small aperture of IUE (see UV1) telescope.

Since scattering by small grains can be considered as nearly isotropic, 
it should also be observed at larger distances from the 
illuminating star.
This is proved not to be the case (UV1):
the scattering optical depth of a nebula varies as $1/\lambda$, not 
as $1/\lambda^4$.

One phenomenum can confine the efficiency of 
scattering to angles close to $0$.
While the scattering optical depth of a low column density 
medium is proportional to the number of scatterers, 
it becomes proportional to the square of the number of scatterers in 
the
case of coherent scattering.
The following sentences, from \citet{bohren}, will best 
resume the idea:

\emph{`Except near the forward direction, there is a random distribution
of phase differences for light scattered by randomly
separated particles in a large collection. 
As we approach the forward direction, however, 
the phase difference approaches 0 regardless of the particle separation. 
Therefore, scattering near the forward direction is coherent. 
If the particles are not identical, the difference in phase between 
light scattered by various pairs of particules, does not, in general, 
vanish in the forward direction, although it is independant of particle separation; 
the phase difference may however, depend on the relative orientation of the two particles. 
It is clear that scattering in or near the forward direction is sufficiently singular to require careful consideration.'}
(\citet{bohren}, p.68)

\citet{jackson}, Sect.9.6, p.417-418, also outlines that coherent scattering 
depends primarely on the exact distribution of scatterers 
in space, except in the forward direction.
For randomly distributed scatterers the scattering is e.g. incoherent 
except for $\varphi$, the angle of scattering, close to 0.

A plane wave disturbed by a particle with size small compared to its' 
wavelength will only be little perturbed and will give rise to a 
secondary wave of amplitude $A$ and a certain phase difference.
A second identical particle will disturb the plane wave in the same 
manner. 
There is no phase difference between the two secondary waves once they 
have reached the observer.
The amplitude of the secondary wave received by the observer is $2A$, 
and the intensity is $4A^2$.
If $N$ identical particules intercept the same plane wave, the 
scattered intensity in direction $\varphi=0$ will be 
$N^2A^2$.

The ratio of coherent to incoherent scattering is 
proportional to the number of small grains, $N$, which of course can be very 
large.
This large difference is the only reason I have found which explains 
the large contamination of the spectrum of reddened stars by scattered light.
\subsection{Variation of the importance of the scattered light with wavelength} 
\label{scaimpo}
The analysis of the spectrum of reddened stars done in UV2 and UV3 
separates three wavelength domains, the extension of which depends on 
the reddening of the star.
These domains are the long and short wavelength spectral regions, or 
small and high optical depth regions, and 
the bump region. 
The two former domains closely coincide with the optical and far-UV regions 
although the frontier is not so clear and is displaced towards the 
optical or towards the far-UV whether the column density of 
interstellar matter is increased or diminished.

The amount of scattered light depends on the proportion of 
extinguished light, hence on the optical depth,
proportional to the column density of interstellar 
matter and increasing as $1/\lambda$.

For very low column density directions, extinction is low. 
The scattered light represents a small (negligible) fraction of the light received 
from the direction of the star in the 
optical and in the UV.
In this case the  spectrum of the star reflects the exact 
extinction law of light by interstellar particles.

The extinguished light rapidly (exponentially) increases with column density.
Scattered light will first appear in the far-UV as an excess on the 
tail of the exponential decrease of the direct starlight (UV2).
Further increase of the column density will see the merging of 
scattered light in the near-UV and to a lesser extent in the optical 
(UV3).
Although scattered light becomes a larger part of the total light received 
from the direction of the star, the effect of linear extinction is to 
diminish the total amount of light (direct+scattered) we receive.
\subsection{The $2200\,\AA$ bump} 
\label{bump}
The low and the large optical depths domains are clearly separated 
since in 
each of them the spectrum is dominated by either the direct or the 
scattered light.
There is no specific reason to question the standard interpretation of 
the $2200\,\rm\AA$ bump, attributed to extinction by a special type of 
grain, but the peculiar position of the bump in between these 
two domains is particular and deserves closer attention.
For small grains there is a phase relation between the scattered and 
the source wave.
If direct and scattered light have  similar weights in the bump region, 
can they interfer and produce an interruption of the scattered light?
\subsection{Angular extent of coherent scattering}
\label{scaext}
There is no phase lag between the waves scattered by two particles
if the particles are on the same line of sight.
When the particles are spread away from the direction of the star a phase relation is maintained 
between the scattered waves as far as they remain within the first 
Fresnel zones.
Coherent scattering extends to the region of space centered on the 
star within which the distance to the observer does not differ by more 
than a few wavelengths.

A difference of a few wavelengths between the distances from the observer 
to points of the Fresnel zones is extremely small compared to astronomical 
distances.
Despite this disproportion, the domain of coherent scattering 
estimated hereafter
includes a region of several hundreds of A.U., which represents 
considerable amount of interstellar matter. 

The occurence of coherent scattering also requires the scatterers to 
be close enough for the source wave to be in phase at their position.

Let $d_0$ be the distance of a background star to the cloud, $h$ the 
distance of a particule in the cloud  to the line of sight of the 
star, $d$ the distance star-particle ($d^2=h^2+d_0^2$), 
$l$ the distance sun-cloud ($h=l\theta$).
$\phi$ will be the diameter of the star.

The spatial extent of the cloud around the direction of the star within which 
starlight can be assimilated to a wavefront
is defined by: $d-d_0\ll \lambda$ and $h\ll \lambda D/\phi$.
Since we must have $h\ll d$, $d\sim d_0$, coherent scattering implies:
$(h/d_0)^2\ll 2\lambda/d_0$ and $h\ll \lambda d_0/\phi$.
The second condition is automatically fullfilled if the first one is 
and if $\phi \ll h$, which is supposed hereafter.

Similarly, viewed from earth, 
the cloud extent over which the phase difference between the light scattered by two particles 
is small satisfies:
$(h/l)^2\ll 2\lambda/l$.

Viewed from earth, the particule is at angular distance $\theta_c$ from 
the direction of the star.
The angular distance from the star within which coherent scattering 
occurs satisfies the two conditions:
\begin{eqnarray}
  \theta_c&\ll& \left(2\frac{\lambda}{l}\frac{d_0}{l}\right)^{0.5}  
    \label{eq:p1} \\
  \theta_c&\ll& \left(2\frac{\lambda}{l}\right)^{0.5}  
    \label{eq:p1b}
\end{eqnarray}¥
If $\lambda=2000\,\rm\AA$ is adopted as an average value for the UV 
wavelength range, and $\theta_c$ in arcsecond:
\begin{eqnarray}
  \theta_c\,(\mathrm{\textquotedblright
})&\ll &
  10^{-7}(\mathrm{\textquotedblright
})\left(\frac{100\mathrm{pc}}{l}\right)^{0.5}
  \left(\frac{d_0}{l}\right)^{0.5}  
    \label{eq:p2}\\
    \theta_c\,(\mathrm{\textquotedblright
})&\ll &
  10^{-7}(\mathrm{\textquotedblright
})\left(\frac{100\mathrm{pc}}{l}\right)^{0.5}
    \label{eq:p2b}
\end{eqnarray}¥
The angle within which coherent scattering 
occurs is much smaller than the resolution accessible to an observation.
\subsection{Column density of the small grains}
\label{gcl}
Let $S=\pi\theta_cl^2$ be the surface of the cloud within which 
coherent scattering occurs, $N$ the column density of the small 
grains, $\sigma$ their cross-section at 
wavelength $\lambda$, $L_\star$ the luminosity of the star at 
wavelength $\lambda$, 
$P_R$ the energy of the scattered light at wavelength $\lambda$ 
received on earth per unit of time and per unit surface, $a$ 
the mean dimension of the small grains.

The power at wavelength $\lambda$ scattered by one small grain and 
received on earth per unit time and unit surface is: 
$\sigma L_\star/(4\pi d_0^2)/(4\pi l^2)$. 
For the $NS$ small grains in $S$ which coherently scatter starlight 
in the direction of the observer:
\begin{equation}
P_R=\frac{\sigma L_\star (NS)^2}{(4\pi d_0 l)^2}
\label{eq:precu}
       \end{equation}¥
The maximum of $P_R$ observed in UV2 was obtained for $E(B-V)\sim 0.1$ 
($A_V\sim 0.3$).
It was reached at $1/\lambda \sim 7\mu\rm m^{-1}$ and is of order 
$15\%$ of the direct light received from the star and corrected for 
reddening: 
\begin{equation}
(P_R)_{max}=0.15\frac{L_\star}{4\pi D^2} \label{eq:pmax}
    \end{equation}
For $P_R$ to equal $(P_R)_{max}$, $N$ must satisfy:
\begin{equation}
\sigma(\frac{D}{d_0})^2\frac{(NS)^2}{4\pi l^2}=0.15
\label{eq:densmax}
    \end{equation}¥
$\sigma$ must be of order $8\pi(2\pi/\lambda)^4a^6$ 
(\citet{vandehulst}, section~6.4).
Equation~\ref{eq:densmax} transforms to:
\begin{equation}
3\pi^3 N \frac{a^3\theta_c^2l}{\lambda^2}\frac{D}{d_0}\sim 0.4
\label{eq:densmax1}
    \end{equation}¥
$D/d_0$ is less than 1.
Use of equation~\ref{eq:p1b} changes 
equation~\ref{eq:densmax1} into:
\begin{equation}
N \gg (10^7\,\mathrm{cm}^{-2}) (\frac{\lambda}{a})^3(\frac{1000\,\mathrm{\AA}}{\lambda})^2
\label{eq:dens1}
\end{equation}¥
The wavelengths for which this inequality applies are close to 
$1000\,\rm\AA$.
From the \citet{bohlin} relation $A_V/N_H \sim 5\,10^{-22}\,\rm mag/cm^2$, 
the hydrogen column density $N_H$ of the 
medium will be $\sim 6\,10^{20}\,\rm cm^{-2}$, for $A_V\sim 0.3$.
Condition~\ref{eq:dens1} takes the two equivalent forms:
\begin{eqnarray}
N &\gg &(10^7\,\mathrm{cm}^{-2}) (\frac{100\,\mathrm{nm}}{a})^3
\label{eq:dens2} \\
\frac{N}{N_H} &\gg &(10^{-13}) (\frac{100\,\mathrm{nm}}{a})^3
\label{eq:dens3}
\end{eqnarray}¥
For exemple if the grains are $1$~nm in size, their column density is 
expected to be larger than $10^{13}$ particle per $\rm cm^2$, and there 
must be more than one small particle per $10^7$ H-atom.
\subsection{Effect of the distance sun-star on the scattered starlight}
\label{scadis}
The distance star-cloud ($d_0$) intervenes only in formula~\ref{eq:p2} 
and will be used only if the star is close or embedded in the 
cloud.
But in most cases $d_0/l$ will be close or larger than 1.
For stars far enough behind the cloud, $\theta_c$ is given by 
relation~\ref{eq:p2b} and the ratio of scattered light to direct 
starlight does not depend on the 
distance of the star.
\section{Scattering by large grains at $\theta=\varphi=0$}
\subsection{Coherence at $\varphi=0$ and large grains}
The size of the particle probably accounts for the absence 
of coherent scattering evidence by large grains in the spectrum of 
reddened stars.
Unless they are identical, large grains will have different 
reflectances and will not give rise to the 
same amplification of the scattered light in the forward direction as 
small grains do.
\subsection{Extension of the $1/\theta^2$ law to $\theta=0$}
My idea when writing UV1 and UV2 was that the $1/\theta^2$ law which is 
sometimes followed by the surface brightness of a nebula implies 
large brightnesses at very small angles and possibly explains the additional 
scattered light in the spectrum of reddened stars.
This idea implies that the $1/\theta^2$ law extends over a large range 
of $\theta $ value, 
from $\theta \sim 10^{-8}\,$'' to a few arcminutes (UV1 and UV2).

According to equation~7 of UV1, the difference of maximum surface 
brightnesses of a nebula, observed in the direction 
of a star, with two different 
apertures, $\theta_1$ and $\theta_2$ (corresponding to solid angles 
$s_1$ and $s_2$) is
$\sim 2\ln(\theta_1/\theta_2)=\ln(s_1/s_2)$ of the flux of the star 
mesured on earth and corrected for extinction.

Hence, if the $1/\theta^2$ law extends to very small angles, we 
expect the absolute reduced spectrum of a star observed behind a cloud 
with the large ($s_1=200\,\rm arcsec^2$) and the small ($s_2=9\,\rm arcsec^2$) 
IUE apertures to differ by $\sim 3\%$ of the spectrum of the star 
corrected for extinction.

In UV2 the absolute reduced spectrum was estimated for a few 
directions (figure~4 to 6 in UV2, figure~2 in UV3).
The far-UV brightness of the scattered light 
ranges from a few $0.1\,\%$ to a $\sim 15\,\%$ of the direct starlight 
corrected for extinction.
A difference of $3\,\%$ of the direct starlight between the light 
scattered in the small and the large apertures of the IUE telescope 
is large enough to be observed.

Since this is not the case, we must conclude that a break of 
the $1/\theta^2$ law occurs at an angle larger than the IUE apertures.
And probably that the contribution of the light scattered by large 
grains is a negligible part of the spectrum of a star.

The final argument which makes unprobable an important 
contribution of light scattered by large grains in the spectrum of a 
star is the scattering cross-section of the large grains.
The scattering optical depth of a nebula observed in the UV
is a linear function of $1/\lambda$ (UV1).
For reasons of continuity, the light scattered by the large grains in 
a beam centered on the star should also be linear in $1/\lambda$.
Since the scattered component in the spectrum of HD46223 
varies as $1/\lambda^4$ (UV3), 
the large grains cannot contribute efficiently to the spectrum 
observed in the direction of the star.
\subsection{Brightness of a cloud due to large grains in the direction of a star}
\begin{figure*}[h]
\resizebox{!}{\columnwidth}{\includegraphics{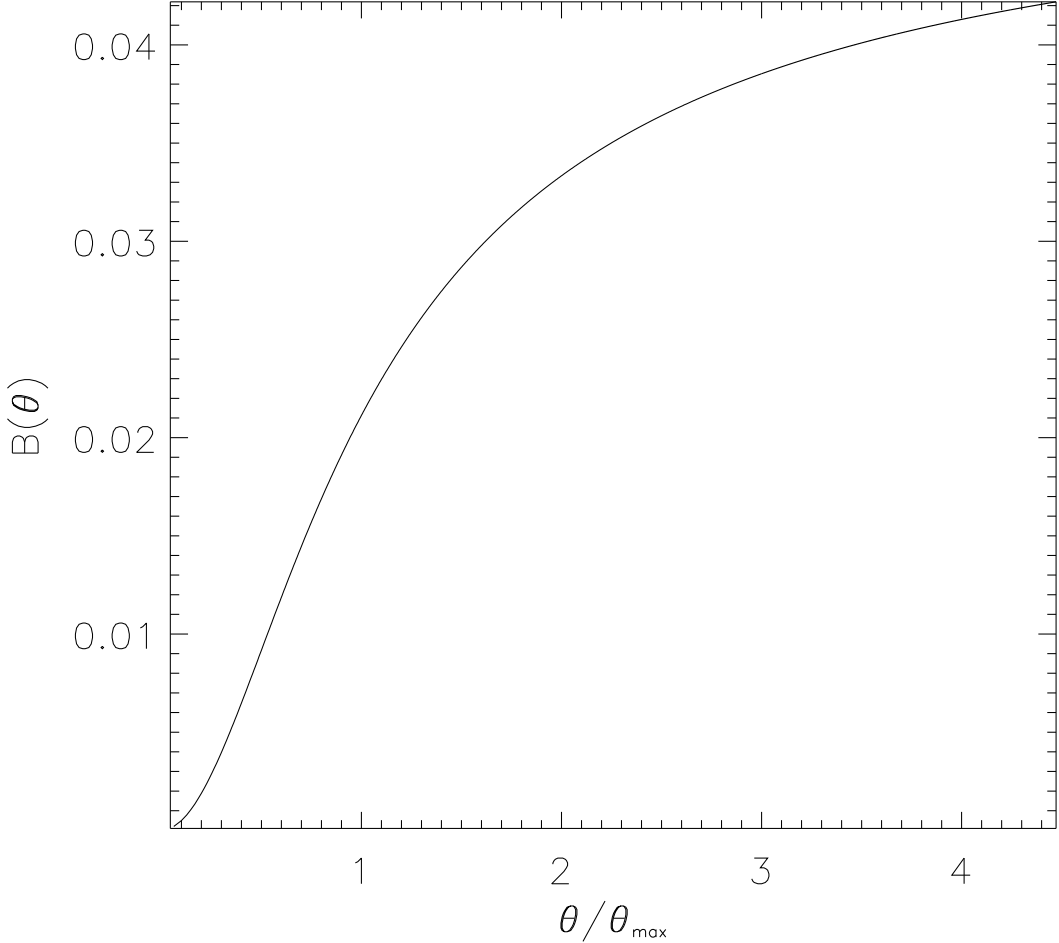}} 
\caption{Large grains' brightness 
for a variable aperture $\theta$ centered on the star. 
The absissa is $\alpha^{0.5}=\varphi/\varphi_{max}=\theta/\theta_{max}$.
} 
\label{fig:sbn}
\end{figure*}
In this section the limits of the $1/\theta^2$ law 
and the maximum surface brightness of the light scattered by large 
grains in a beam centered on a star are investigated.

As in UV1, let $R$ be the ratio of the surface brightness of a cloud, 
observed at distance $\theta$ to a star,
to the flux of the star measured on earth and corrected for extinction.
$g$ will be the phase function of the grains, $D$, $d_0$, and $\varphi$ 
($\sin \varphi =D/d\sin\theta$) were defined previously and 
correspond to the notations of figure~4 of UV1.
$S_0(\tau,\varphi)$ will be the surface brightness of a cloud of 
optical depth $\tau$ in direction $\varphi$ and for a source radiation 
field of $1$ unit (power/unit surface/wavelength) at the cloud 
location.

Since the maximum possible surface brightness of the cloud needs to be 
estimated, $\tau$ is supposed to be optimized so that $S_0$ is maximum 
for each value of $\varphi$.
$S_0$ depends on $\varphi$ only and is proportional to $g$: 
$S_0=\alpha_0 g(\varphi)$.
Then:
\begin{equation}
R=S_0\sin^2\varphi/\sin^2 \theta
=\alpha_0 g(\varphi)\sin^2\varphi/\sin^2 \theta
	\label{eq:r0}
\end{equation}¥
$S_0\sin^2\varphi$ is $0$ at $\varphi=0$. 
For forward scattering grains it must also be small at $\varphi=\pi/2$.
In between, $S_0\sin^2\varphi$ will have a maximum at $\varphi_{max}$.
The more scattering is oriented in the forward direction, the smaller $\varphi_{max}$ is.

$S_0\sin^2\varphi\sim \alpha_0 g(\varphi) \varphi^2$ can be developped 
into powers of $(\varphi-\varphi_{max})^2$.
The first term of the development is the zero order term and the second term depends 
on $(\varphi-\varphi_{max})^2$.
For $\varphi$ close to $\varphi_{max}$ $R\theta^2=S_0\varphi^2$ is 
constant.
It is around $\varphi_{max}$ that the $1/\theta^2$ law will best hold.

The nebulae for which the $1/\theta^2$ law applies can reasonably 
be thought to be observed at $\theta$ close to $\theta_{max}= \varphi_{max}d/D$.
Since the nebulae observed with the IUE telescope were at angular 
distances ranging from a few tens of arcsecond to a few arcminute, 
$\varphi_{max}=\theta_{max}D/d>\theta_{max}$ is probably larger 
than a few tens of arcseconds.

After a plateau around $\theta_{max}$, $R\theta^2=S_0\varphi^2$ will quickly level 
off to 0 with $\theta \rightarrow 0$.
Hence, the light scattered by large grains in a small solid angle centered 
on the star is far less than the light scattered in the same solid angle 
around $\theta_{max}$.
It is a negligible (small compared to $c=(R\theta^2)_{max}\sim$ a few $10^{-3}$, UV1) 
part of the flux of the star.

The brightness of the nebula observed with a beam centered on the star 
can be estimated with Henyey-Greenstein's function as the phase function of 
the grains.
Henyey-Greenstein function can be developed in the small angles 
approximation.
Equation~\ref{eq:r0} becomes:
\begin{eqnarray}
R&=&\alpha_0 g(\varphi)\varphi^2/ \theta^2 \nonumber \\
&=& \alpha_0
  \frac{1}{\pi\varphi_{max}^2(1+2\alpha)^{1.5}}(D/d)^2
    \label{eq:maxred1}\\
    & \,=\, & \alpha_0 \frac{\alpha}{\pi(1+2\alpha)^{1.5}}
    \frac{1}{\theta^2},
    \label{eq:maxred2} 
\end{eqnarray}¥
with $\alpha=(\varphi/\varphi_{max})^2$, $\varphi_{max}\sim 1.47(1-g_0)$.
$g_0$, the asymetry parameter of the phase function, was supposed to 
be close to $1$ corresponding to a strong forward scattering phase 
function.

According to equation~\ref{eq:maxred2}, the maximum of $R\theta^2$ 
is $\alpha_0/(3^{1.5}\pi)=0.06\alpha_0$, obtained for $\alpha=1$.
Since $c=R\theta^2$ was estimated to be $\sim 3\,10^{-3}$,
we deduce: $\alpha_0 \sim 5\,10^{-2}$. 

For a star behind the cloud, $d\sim d_0$ is nearly constant.
The ratio (brightness of the cloud)/(unreddened flux of the star) is:
\begin{eqnarray}
  B(\theta)= \int_0^\theta 2\pi\theta R{\rm d}\theta  & = &
    \int_0^\theta 
    \alpha_0\frac{2\pi\theta{\rm d\theta}}{\pi\varphi_{max}^2(1+2\alpha)^{1.5}}
    \left(\frac{D}{d_0}\right)^2 \nonumber \\
     &=& \alpha_0\int_0^\alpha
     \frac{1}{(1+2\alpha)^{1.5}}
     {\rm d}\alpha \nonumber\\
     & = & \alpha_0(1-\frac{1}{(1+2\alpha)^{0.5}}) \label{eq:ba}  \\
  &= &      
     \alpha_0\left(1-\frac{1}{\left(1+2\left(     
      \frac{D}{d_0}\frac{\theta}{\varphi_{max}}\right)^2\right)^{0.5}}\right)
\label{eq:bt}
 \end{eqnarray}
Figure~\ref{fig:sbn} plots $B(\theta)$ (for $\alpha_0 = 5\,10^{-2}$) as a 
function of $\theta/\theta_{max}=(D/d_0)\theta/\varphi_{max}$.
$B(\theta)$ increases with $\theta$ up to a maximum brightness $B_{max}=\alpha_0$, $\sim 
5\% $ of the direct starlight.

$B(\theta)$ approaches $B_{max}$ when $(D/d_0) (\theta/\varphi_{max})$ 
is more than a few units.
If $\theta$ is a few arcsecond (IUE apertures), with $\varphi_{max}$ 
larger than a few tens of arcsecond, $\theta/\varphi_{max}$ 
is less than $0.1$.
$B(\theta)$ is significant only if $D/d_0$ is large, which needs the star 
to be close to the cloud.

For a cloud at a few hundred parsec from the sun most stars observed 
behind the cloud can be supposed to be far enough and $D/d_0\sim 1$.
Equation~\ref{eq:bt} transforms into:
\begin{equation}
    B(\theta)=\alpha_0 
    \left(\frac{D}{d_0}\frac{\theta}{\varphi_{max}}\right)^2
   =\alpha_0 
    \left(\left(1+l/D\right)\frac{\theta}{\varphi_{max}}\right)^2
    \label{eq:bmin}
\end{equation}¥
For these stars observed with the IUE telescope $B(\theta)$ is 
negligible, a few $10^{-4}$ at most.

It follows that the light scattered by large grains in front of the 
star will in general be a negligible part of the direct starlight.
Only when the star is embedded in the nebula can it represent a few 
per cent of the direct starlight, still insufficient to account for the $15\% $
of scattered light observed in the spectrum of some reddened stars.
\section{UV observations of nebulae}
UV spectra of nebulae (UV1) point to large grains, with identical 
properties (optical depth linear in $1/\lambda$, strong forward 
scattering phase function)
as in the optical, as the carriers of the scattering.
It is then tempting to assume continuity of the scattering properties of the 
large grains from the optical to the UV.
A contrario the standard theory of extinction clearly separates optical and UV 
wavelengths and attributes the inter-action light-particles to different 
kind of interstellar grains in each of the wavelength domains.
The following sections pursue the implications of the observations of 
nebulae in the UV for the standard theory and in the case of an 
identical extinction law in the optical and in the UV.
\subsection{UV observations of nebulae and the standard interpretation 
of the extinction curve}
The three component model associated to the standard interpretation (see 
 \citet{desert} for the decomposition of the extinction curve, or figure~3 in 
the more recent review of \citet{greenberg}) implies a nearly constant extinction 
in the UV of the large grains 
responsible for the light scattered by a nebula.
The rupture between the UV and the optical properties of the large 
grains is strong and occurs between  $1/\lambda=2\,\mu\rm m^{-1}$ 
and $1/\lambda=3\,\mu\rm m^{-1}$.
Above $1/\lambda=3\,\mu\rm m^{-1}$, the UV extinction coefficient $A_\lambda$ 
due to large grains is nearly constant in the UV and close to $A_V$: 
$A_\lambda \sim A_V \sim 3E(B-V)$ for $1/\lambda>3\,\mu\rm m^{-1}$.
The remaining UV extinction is due to small grains.
Starlight in the bump spectral range is extinguished 
by small grains with $A_{\lambda_b}\sim 2E(B-V)$ (figure~3 in 
\citet{greenberg}). 

UV observations of nebulae (UV1) prove that $S_{neb}/F_\star$, the ratio of 
the brightness of the nebula to the flux of the illuminating star, 
is linear in $1/\lambda$ and varies by factors larger than $2$ 
between $1/\lambda=3\,\mu\rm m^{-1}$ and $1/\lambda=8\,\mu\rm m^{-1}$.

To reconcile the standard theory with these observations we must 
admit a singular combination of circumstances which implies:
\begin{itemize}
    \item  the scattering optical depth in the UV of the small grains is 
   either negligible or varies as $1/\lambda$,

    \item  the albedo of the large grains is linear in $1/\lambda$

\end{itemize}¥
Hence we would have to conceive that the large grains have an 
extinction cross section which varies as $1/\lambda$ in the optical 
and constant in the UV and an albedo constant in the optical and 
varying as $1/\lambda$ in the UV.

For nebulae such as the $17\tau$ nebula or the Orion nebula which do 
not have a bump (UV1), the problem is even harder.
Two cases are conceivable. 
Either the nebula is different from and of higher column density 
than the medium in front of the star, in which case a bump should be 
observed in the spectrum of the nebula.
Or the nebula is of very low column density, $A_\lambda \sim  0.15$ 
in the UV, as it is in the directions of HD23302 and of HD37742 ($E(B-V)\sim 
0.05$), in which case it is difficult to understand the large amount of scattered light by 
the nebula and its' variations with wavelength.
\subsection{Alternative interpretation of the UV spectrum of the 
nebulae}
In clear opposition with the standard interpretation of the 
extinction curve UV1 and UV2 have emphasized the possibility of 
continuous properties of the large grains in the optical and in the UV.
This continuity contredicts the near-UV rupture imposed in the standard 
theory and recalled in the preceding section.
$A_\lambda$ regularly increases in the UV as $1/\lambda$.

There is no bump in the spectrum of $17\tau$ while there is 
an interstellar cloud between the star and us corresponding to the 
reddening of $17\tau$: $E(B-V)\sim 0.05$.
If the nebula observed $18''$ away from $17\tau$ belongs to the same 
cloud, a reasonnable hypothesis, its' optical depth $\tau_\lambda\sim 
2E(B-V)/\lambda$ (see UV2), will vary from $0.3$ to $0.8$ for the UV 
wavelength range of IUE observations.
Which means that appreciable scattered light is expected from the 
nebula, as it is observed.

In a similar way, the cloud in front of HD~37742 ($E(B-V)\sim 0.07$) may not have enough 
column density to produce an observable bump in the spectrum of the 
star and in the spectrum of the nebula but will be bright at UV 
wavelengths since its' UV optical depth varies from $0.4$ to $1$.

Generally speaking suppose a nebula is observed close enough to the 
illuminating star so that it is similar to the medium in front of the 
star.
Coherent scattering in the forward direction ($\varphi\sim 0$) 
is created by the radiation field entering the nebula in the 
same way as it is created in the direction of the observer.
This scattered light cannot be neglected since it can be of the 
same order of magnitude as the direct starlight.
It is added to the source radiation field 
and will in turn be scattered in the direction of the observer as the direct starlight is.
The brightness of the nebula is proportional to the sum 
of the source radiation field and of the coherently scattered light, 
hence to what is observed in the direction of the star.

The surface brightness to star flux ratio departs 
from the linear relation in $1/\lambda$ only if the nebula is 
different from the medium in front of the star.
This does not seem to be the case for IUE observations.
\section{Conclusion}
In this paper there are three main questions I have tried to answer.
How can small particles which isotropically scatter starlight
produce the large amount of scattered light in the forward direction
necessary to explain the UV spectrum of the stars?
Why is scattering at $\theta=0$ by large forward scattering particles negligible 
compared to the direct starlight?
Why is the spectrum of a nebula the exact product of the spectrum of 
the star and a linear function of $1/\lambda$?
And more specifically why does the UV spectrum of a nebula have a 
bump at $2200\,\rm\AA$ if and only if the spectrum of the illuminating star also has a 
bump?

The only possibility for randomly distributed particles to
efficiently scatter starlight in the forward 
direction is coherent scattering.
Then, the scattering optical depth varies as 
$N^2$, the square of the number of scatterers, instead of $N$.
This will explain the importance of the scattered light observed in 
the spectra of reddened stars.

Coherent scattering concerns more likely small grains, with nearly 
identical reflectances, than large grains.
Coherent scattering is limited to a small solid angle around the 
star, much smaller than the resolution accessible to an observation.
Although small this angle encompasses enough matter to 
provide the necessary amount of scattered light with reasonably small column densities of 
small grains.

The presence of a substantial amount of coherently scattered light in 
the spectrum of reddened stars can be derived from two distinct points of 
vue.
On one hand spectrum of reddened stars contamination by scattered 
light is explained well by the presence of grains small compared to UV 
wavelengths.
On the other hand, based on this paper's calculations, if very small 
grains do exist in interstellar space, they will coherently scatter 
light in the forward direction and therefore contribute substantially to the 
spectrum of the star.

The  $2200\,\rm\AA$ bump region deserves attention.
The standard interpretation of the bump states it comes from extinction due 
to a special class of interstellar grains.
It remains the most plausible one but
UV2 and UV3 point out the possibility that the bump can be extinction 
of only the scattered light. 
It is also in this region that scattered and direct light have similar weights. 
Since direct and scattered light can interfer, the hypothesis was 
formulated that the bump could be the result of a destructive interference.

Scattering by large particles gives a maximum of $R\theta^2$ when 
the angle of scattering is close to a certain angle $\varphi_{max}>0$.
In the case of interstellar grains at UV wavelengths $\varphi_{max}$ 
is probably of a few tens of arcseconds or more.
If $\theta$ is decreased to $0$, when $\varphi$ becomes smaller 
than $\varphi_{max}$, $R\theta^2$, and the brightness of the nebula in 
the direction of the star observed with resolution $\theta$,  also tend to $0$.
The brightness of the nebula, due to the large grains, in the direction 
of the star, can be estimated with Henyey Greenstein phase function. 
It should not exceed a few $10^{-4}$, a low $\% $ in the most 
favorable of cases, of the flux of the star measured at the position of 
the observer and corrected for extinction.

A bump at $2200\,\rm\AA$ is created in the light scattered by a nebula 
in the same way as it is in the direction of the star.
The light which enters in the nebula is coherently scattered in the 
forward direction.
Both direct starlight and coherent scattered light are scattered in 
the direction of the observer.
If the medium on the line of sight of the star and in the direction 
where the nebula is observed do not differ appreciably, proportional bumps 
are created.
{}
\end{document}